# LANGUAGE TIME SERIES ANALYSIS

Kosmas Kosmidis, Alkiviadis Kalampokis and Panos Argyrakis

Department of Physics, University of Thessaloniki, 54124 Thessaloniki, Greece

**Abstract**

We use the Detrended Fluctuation Analysis (DFA) and the Grassberger-Proccacia analysis (GP) methods in order to study language characteristics. Despite that we construct our signals using only word lengths or word frequencies, excluding in this way huge amount of information from language, the application of Grassberger-Proccacia (GP) analysis indicates that linguistic signals may be considered as the manifestation of a complex system of high dimensionality, different from random signals or systems of low dimensionality such as the earth climate. The DFA method is additionally able to distinguish a natural language signal from a computer code signal. This last result may be useful in the field of cryptography.

# 1. INTRODUCTION

Human language has recently attracted the attention of the physical scientists. Following the advances in the theory and understanding of complex systems, it was recently realised that human language is a new emerging field for the application of methods from the physical sciences in order to achieve a deeper understanding of linguistic complexity. Important work in the field of the mathematical modelling of language and in the field of language simulations has recently been done by several groups [1-13]. There is also renewed interest in the task of discovering and explaining structural properties of the languages, such as the Zipf law [14-16] which actually deals with the probability distribution of words in spoken languages. Of course, the understanding of the complexity associated with language is not an easy task. We have to use all kinds of our mathematical tools in order to gain understanding of the system we study. One of these tools is Time Series Analysis [17]. Time series analysis plays a key role in physical sciences. Our goal is to extract information from signals that are related to real-world phenomena. Analyzing such signals allow us to achieve better understanding of the underlying physical phenomena. The methods of analyzing signals are wide spread and range from classical Fourier analysis to various types of linear time-frequency transforms, model-based and non-linear approaches.

A particularly interesting characteristic in time series associated with several physical processes is the presence of long-range correlations. Some interesting examples include DNA sequences [18-22], weather records [23] and heart rate sequences [24-30]. The common feature of all these diverse systems is that the long-range correlations decay by a power law, where a characteristic scale is absent. These findings are useful, e.g., in DNA for distinguishing between coding and noncoding sequences [22], in atmospheric science for testing state-of-the-art climate models, etc.

Long range correlations may be detected using a method called Detrended Fluctuations Analysis [31,32], which we will presented in section 3.

Moreover, when we study time series and we do not rely on any particular model we are interested in getting an insight of the dynamics of the system solely from the knowledge of the time series alone. In such cases a method derived by Grassberger and Proccacia [33-35] has been proven particularly useful. This method has been applied to analyze the dynamics of climatic evolution [35 and references therein], neural network activity [36], or electric activity of semiconducting circuits [37,38]. Details on the method are presented in section 4.

## 2. MAPPING DOCUMENTS TO TIME SERIES

The main problem one has to deal with before applying the analysis methods is the following: Given a document written in natural language how can one transform it in a time series and then analyze it? Although, at first, time series and natural language documents seem to be irrelevant we will present two ways to construct time series from documents:

i.  Take a document of $N$ words. Count the length $l$ (number of letters) of each word. The role of time is played by the position of the word in the document i.e. the first word is considered to be emitted at time $t=1$, the second at time $t=2$ etc. We map the word length to this time and thus a time series $l(t)$ is constructed.. Henceforth, we will refer to such time series as "Length time series" (LTS).

ii.  Take a document of $N$ words. Count the frequency $f$ of appearance of each word in the document. Again the role of time is played by the position of the word in the document. We map the word frequency to this time and thus a time series

*f(t)* is constructed. Henceforth, we will refer to such time series as "Frequency time series" (FTS).

Obviously there is a large number of ways to map a document to a time series but in the present study we deal with the above two as there is also a physical meaning in the mapping. The length of the word is associated with speaker effort, meaning that the longer the word the higher the effort required to pronounce it. The frequency of the word is also associated with the hearer effort as frequently used words require less effort to be understood from the hearer.

Linguistic series have been studied in the past [39-41] but dealt human writing at letter level and not at word level. Human writings at word level were studied by Montemurro et al [42]. They use a "frequency mapping" similar –but not identical- to the one described above but they use a different method for their analysis. So their results have to be considered as complementary to ours.

## 3. DETRENDED FLUCTUATIONS ANALYSIS

The DFA estimates a scaling exponent from the behaviour of the average fluctuation of a random variable around its local trend. The method can be summarized as follows. For a time series $u_t$, $t = 1,2, \ldots, N$, first the integrated time series $Y$ is obtained:

$$Y(i) = \sum_{t=1}^{i} [u_t - <u>] \tag{1}$$

where $<u>$ is the sample mean.

In the second step, we divide $Y(i)$ into $N_s \equiv [N/s]$ non-overlapping segments of equal length $s$. Since the record length $N$ need not be a multiple of the considered time scale $s$, a short part at the end of the profile will remain in most cases. In order not to disregard this part of the record, the same procedure is repeated starting from the other end of the record. Thus, $2N_s$ segments are altogether obtained.

In the third step, we calculate the local trend for each segment $v$ by a least-square fit of the data. We denote the detrended time series for segment duration s by $Y_s(i)$. It is the difference between the original time series and the fits:

$$Y_s(i) = Y(i) - p_v(i) \qquad (2)$$

where $p_v(i)$ is the fitting polynomial in the $v_{th}$ segment. If quadratic polynomials are used in the fitting procedure, the method is called quadratic DFA (DFA2). Linear, cubic, or higher order polynomials can also be used (DFA1, DFA3, and higher order DFA).

In the fourth step, we calculate the variance for each of the $2N_s$ segments

$$F_s^2(v) = <Y_s^2(i)> = \frac{1}{s}\sum_{i=1}^{s} Y_s^2[(v-1)s+i] \qquad (3)$$

for the detrended time series $Y_s(i)$ by averaging over all data points i in the $v_{th}$ segment.

Finally, we average over all segments and take the square root to obtain the DFA "Fluctuation function":

$$F(s) = [\frac{1}{2N_s}\sum_{v=1}^{2N_s} F_s^2(v)]^{1/2} \qquad (4)$$

If only short range correlations (or no correlations) exist in the time series, then the "Fluctuation function" will have the statistical properties of a random walk. Thus, we expect $F(s) \sim s^\alpha$ with $\alpha=1/2$, while in the presence of long range correlations $\alpha \neq 1/2$

## 4. GRASSBERGER –PROCCACIA ANALYSIS

At first sight, a time series of a single variable appears to provide a limited amount of information. We usually think that such a series is restricted to a one-dimensional view of a system, which, in reality, contains a large number of independent variables. It has been shown [33-35], however, that a time series bears the marks of all other variables participating in the dynamics of the system and thus we are able to "reconstruct" the systems phase space from such a series of one- dimensional observations. When applying the Grassberger-Proccacia (GP) method to a time series we want to find the answer to the following questions:

i. Can the salient features of the system be viewed as the manifestation of a deterministic dynamics, or do they contain an irreducible stochastic element? Is it possible to identify an attractor in the system phase space from a given time series?

ii. If the attractor exists, what is its dimensionality $d$?

iii. What is the minimal dimensionality, $n$, of the phase space within which the above attractor is embedded? This defines the minimum number of variables that must be considered in the description of the underlying system.

This is done as follows. We consider a time series. Let us call this signal $x_0(t)$. We would like to reconstruct the dynamics of the system solely on our knowledge of $x_0(t)$. We consider the phase space spanned by the variables $k=0, 1, 2. . . n-1$, where $k$ are several variables that take part in the dynamics of the system. For our problem these are the parameters related with language and we do not know which or how many they are. At a given time a state of the system is a point in phase space, while a sequence of states in time gives a trajectory. If the dynamics of the system obey some dissipative deterministic laws, then the trajectories converge to an attractor. We thus form this attractor from the $x_0(t)$ series, by successively shifting the original time series by the same amount in time $\Delta t=\tau$, and forming $n$ such series as

$$\begin{aligned}
x_0 &: x_0(t_1), x_0(t_2), x_0(t_N), \\
x_1 &: x_0(t_1+\tau), x_0(t_2+\tau), x_0(t_N+\tau), \\
x_2 &: x_0(t_1+2\tau), x_0(t_2+2\tau), x_0(t_N+2\tau), \\
x_{n-1} &: x_0(t_1+(n-1)\tau), x_0(t_2+(n-1)\tau), x_0(t_N+(n-1)\tau)
\end{aligned} \quad (5)$$

These variables are expected to be linearly independent if the $\tau$ shift is properly chosen. We chose several different $\tau$ values, but in the subsequent calculations we use $\tau=500$ time units. Notice that $x_0(t)$ is a vector made of the set of points, as given in Eq.5. A general notation for it is $x_i$. We now choose a reference point in $x_i$ and compute all the distances $|x_i - x_j|$ from the ($N$-1) remaining points. This way we get the total of all points $x_i$ in phase space. Doing this for all $i$ we get

$$C(l) = \frac{1}{N^2} \sum_{i,j=1, j \neq i}^{N} \Theta(l - |x_i - x_j|) \qquad (6)$$

Where $\Theta$ is the Heavyside step function, $\Theta(x)=0$, if $x<0$ and $\Theta(x)=1$ if $x>0$. $C(l)$ is the correlation integral of the attractor, since it shows how a point in the vector $x_i$ affects the positions of other points. Thus if the attractor is a $d$-dimensional manifold, then we expect $C \sim l^d$, with its dimensionality given by the exponent $d$.

## 5. RESULTS AND DISCUSSION

Both English and Greek texts were used for this research. Specifically the English texts were the "War of the Worlds" by H. G. Wells, "The Mysterious Affair at Styles" by Agatha Christie and "A Christmas Carol" by Charles Dickens. For the Greek language the translation of "Sangharakshita, Vision and Transformation" was used. All of the above texts were found using Project Gutenberg (www.gutenberg.net). The Greek corpus was complimented with extracts from publications of the Greek newspaper "Ta Nea" for the years 1997-2003.

In Figure 1 we present an LTS signal (a) and an FTS signal (b). This series were constructed from a document written in Greek, taken from the Greek newspaper "Ta Nea". The total number of words used for the analysis is 36221 words, but only a small part (100 words) of the series is shown for clarity.

In Figure 2a we present the results of third order detrended fluctuation analysis. In this method we use a $3^{rd}$ degree polynomial in order to perform the detrending. The resulting function $F_d$ follows a power law but exhibits a cross over at about $s \sim 200$. The initial part has an exponent n =0.46 while the remaining part has an exponent of 0.61. In figure 2b we present the same analysis for the shuffled series. Shuffling should destroy long range correlations leading to an exponent n=0.50. In our data the exponent is found to be 0.52 for the shuffled data, as expected. In our example the linguistic signal appears to be slightly anticorrelated in short time scales. The slope however is rather close to the value 0.5 which indicates absence of correlations. Thus, we have to be rather careful with the above conclusion.

Anticorrelated behavior practically means that a word of short length has a higher probability to be followed by a long word and vice versa. In Greek language short words (usually articles) are often followed by longer words (nouns, adjectives etc) so this anticorrelated behavior is somehow anticipated. This is an indication that the above method may detect the presence of syntactic structure in a linguistic document without knowing any more details of the language.

Moreover the existence of a crossover from an early almost uncorrelated behavior to a long range correlated one, is a characteristic of a signal composed of different patches [25]. Moreover, a location of the crossover is approximately the same as the size of the patches. Here, the crossover is found at s~200. This is roughly the size of a paragraph! This indicates that the DFA method is capable of detecting that the paragraph structure of the document. See the marked difference in the results derived from computer code

In figure 3a we present the same type of analysis for an English language document namely "Christmas Carols" by Dickens. English grammar is somehow different than the Greek. Syntax is not so restricting and articles are not used as often as in Greek documents. Here the resulting slope of the straight line in figure 3a is 0.48 much closer to that of an uncorrelated time series. In figure 3b we perform the same analysis for the frequency signal. The calculated slope is equal to 0.49, almost identical to that of an uncorrelated signal.

The situation is however different if instead of a spoken language we study a computer language. A spoken language is a means of communication between human beings while a computer language is a means of communication between humans and computer machines. We have constructed an LTS signal using the Linux Kernel which is written in C. It must be noted that in computer programs (as the Linux Kernel) both programming language instructions and variable names are used. The instructions are few in number and are language specific, whereas the variable names are chosen by the programmer. In the present work no distinction was made between the two, as the whole program was treated as a means of communication between the programmer and the computer.

In Figure 4a we present the signal and in Figure 4b we present the results of DFA3 analysis. The slope of the straight line is 0.64 indicating the presence of long range correlations. Performing the same analysis on the FTS signal of the Linux Kernel we obtain a straight line with slope 0.55 (data not shown).

Therefore we conclude that there is a marked difference between human and computer languages. Human language LTS and FTS signals are probably non-correlated or slightly anticorraleted whilst computer language LTS and FTS signals exhibit long range correlations.

In Figure 5 we present the results of our Grassberger – Proccacia analysis for LTS signals. Figure 5a shows the Correlation integral $C(r)$ as a function of $r$ for several embedding dimensions $n$ of the phase space for the LTS signal of a Greek language document. Notice the straight line segments of the double logarithmic plot. We use these parts in order to estimate the exponent $d$ and determine the scaling behavior of the correlation integral. Then we plot in Figure 5b the obtained $d$ values as a function of the embedding dimension $n$. For a white noise signal we expect to see a straight line with slope equal to one, while for a low dimensional chaotic system we expect a saturation to some value of $n$ which with allow us to determine the minimal dimensionality of the phase space needed to embed the attractor and consequently to determine how many independent variables we have to consider in describing a linguistic system. What we see is that there is no saturation and also that the slope of the line is 0.69 i.e. rather less than one. Is this an indication that we are dealing with a complex system of high dimensionality? If so, the dimension of the phase space of this system is high, possibly infinite. Thus, there is a marked difference between for example climatic records and LTF signals. DFA analysis of climatic records has revealed that they are strongly long range correlated [42] while a Grassberger-Proccacia analysis has shown that they may be seen as the manifestation of a low-dimensional chaotic system (the embedding dimension was found to be no more than 5)[35]. For Language LTS signals we see that signals that are not so strongly long range correlated, but the dynamics of the system involve a large amount of parameters! Note, however, that when a complex system has a high dimensional phase space it does not mean that it is mathematically intractable. For example, a delay differential equation is in fact an infinite dimensional system [34]. We can, nevertheless, study it numerically and obtain useful results.

In order to investigate the relation between the DFA analysis and the Grassberger-Proccacia method, we have performed a Grassberger-Proccacia (GP) analysis on the language LTS after shuffling the word order. Shuffling will destroy the correlation between words but will maintain the probability distribution of the word lengths. The resulting slope for the shuffled series (Figure 5b, triangles) is practically identical to that of the original LTS. This is probably due to the fact that Grassberger-Proccacia (GP) method is giving information about the geometric structure of the phase space attractor, but it is not sensitive to the particular sequence with which the points on this attractor are visited!

As a finishing remark, we would like to point to a possible application of the DFA method. It seems that the method may be used to distinguish between natural language and computer code. Suppose that we have a encrypted sequence of words. We may construct an LTS signal from this sequence and then apply the DFA method in order to decide whether this sequence is random, natural language or computer code. From our analysis seems that the DFA method may be useful in some aspects at the field of cryptography.

**ACKNOWLEDGEMENTS**

The authors want to thank Prof. Armin Bunde and Prof. Shlomo Havlin for useful discussions. This work was supported by the Greek Ministry of Education through the PYTHAGORAS project.

**REFERENCES**

1. M. Nowak and D. Krakauer, Proc. Natl. Acad. Sci. USA, **96**, (1999),8028.

2. D. Abrams, S. Strogatz, Nature **424**, (2003),900

3. C. Schulze, D. Stauffer, Int. J. Mod. Phys. C, **16**, (2005), 718 and AIP Conference proceedings 119, 49 (2005) (8th Granada Seminar)

4. C. Schulze, D. Stauffer, Phys. of Life Rev. **2**, (2005), 89

5. K. Kosmidis, J.M.Halley, P.Argyrakis, Physica A **353**, (2005), 595

6. K.Kosmidis, A.Kalampokis, P.Argyrakis, Physica A (to be published) = physics/051019

7. J. Mira, A. Paredes, Europhys. Lett. **69**, (2005) 1031

8. V. Schwammle, Int. J. Mod. Phys. C, **16**, (2005) issue 10, 1519 = physics/ 0503238

9. V. Schwammle, Int. J. Mod. Phys. C, **17**, (2006) issue 3= physics/ 0509018

10. T. Tesileanu, H. Meyer-Ortmanns, Int. J. Mod. Phys. C, **17**, (2006) issue 3= physics/ 0508229.

11. M. Patriarca, T. Leppapen, Physica A, **338,**(2004)**,** 296

12. V.M. de Oliveira et al., Physica A (to be published) = physics/0505197

13. V.M. de Oliveira et al., Physica A (to be published) = physics/0510249

14. Zipf, G.K. *Human behavior and the principle of least effort: an introduction to Human Ecology* (Addison-Wesley, Cambridge, MA, 1949)

15. S. Havlin, Physica A **216**, 148-150 (1995)

16. R. Cancho and R. Sole, Proc. Natl. Acad. Sci. USA, **100**, (2003), 788

17. D.S.G Pollock, *Time series analysis, Signal Processes and Applications* (Academic Press, London,1999)

18. C.-K. Peng et al., Nature (London) **356**, 168 (1992)

19. R. F. Voss, Phys. Rev. Lett. **68**, 3805 (1992)

20. S.V. Buldyrev et al., Phys. Rev. Lett**. 71**, 1776 (1993)

21. A. Arneodo, E. Bacry, P.V. Graves, and J. F. Muzy, Phys. Rev. Lett. **74**, 3293 (1995).

22. R.N Mantegna et al., Phys. Rev. Lett **73**, 3169 (1994)

23. S.V. Buldyrev et al., in Fractals in Science, edited by A. Bunde and S. Havlin (Springer, Berlin, 1994).

24. E. Koscielny-Bunde et al., Phys. Rev. Lett. **81**, 729 (1998).

25. C.-K. Peng et al., Phys. Rev. Lett. **70**, 1343 (1993)

26. C.-K. Peng et al., Chaos **5**, 82 (1995)

27. C.-K. Peng et al., Physica (Amsterdam) **249A**, 491 (1998)

28. S. Thurner, M.C. Feurstein, and M. C. Teich, Phys. Rev.Lett. **80**, 1544 (1998)

29. L. A. Amaral, A. L. Goldberger, P. Ch. Ivanov, and H. E. Stanley, Phys. Rev. Lett. **81**, 2388 (1998)

30. P. Ch. Ivanov et al., Nature (London) **399**, 461(1999).

31. C.K. Peng at al, Phys. Rev E,**49(2),** (1994),1685 -1689.

32. J. Kantelhardt, E. Koscielny –Bunde, H. Rego, S. Havlin, A. Bunde, Physica A **295,** 441 (2001)

33. P. Grassberger, I. Procaccia, Phys. Rev. Lett. **50**, 346 (1983)

34. P. Grassberger, I. Procaccia, Physica D, **9**, 189 (1983)

35. G. Nicolis, Ilia Prigogine, *Exploring Complexity,* Freeman and Co, New York, 1989.

36. C. Kotsavasiloglou, A. Kalampokis, P. Argyrakis, S.Balogiannis, Phys. Rev. E, **56(4)**, 4489 (1997)

37. Ch. Karakotsou, A. N. Anagnostopoulos, Physica D **93,** 157 (1996)

38. Ch. L. Koliopanos, I. M. Kyprianidis, I. N. Stouboulos, A. N. Anagnostopoulos, L. Magafas, Chaos Soliton Fract **16**, 173 (2003).

39. A. Schenkel, J.Zhang, Y-C. Zhang, Fractals **1**, 47 (1993)

40. R. F. Voss, Fractals **2**, 1 (1994).

41. M. Amit, Y. Shemerler, E. Eisenberg, M. Abraham, N. Shnerb, Fractals **2**, 1, 7 (1994)

42. M. Montemurro, P.Pury, Fractals **10**, 4, 451 (2002)

43. A. Bunde, J.F. Eichner, J.W. Kantelhardt, S. Havlin, Phys. Rev. Lett. **94**, 048701 (2005)

# FIGURE CAPTIONS

1. (a)Word length versus time. (b) Word Frequency versus time. This series was constructed from a document written in Greek. Only a part of the series is shown (100 words). The total number of words used for the analysis is 36221 words.

2. (a) Plot of the DFA3 "Fluctuation function" $F(s)$ vs. $s$ of the "length time series" presented in Figure 1. A power law behavior is observed and the slope of the straight line is equal to 0.45. The slope change for values of s greater than 200 is probably due to (periodic) trends not eliminated from the DFA3 procedure.(b) Same plot for the shuffled "length time series" presented in Figure 1. A power law behavior is observed and the slope of the straight line is equal to 0.52

3. (a) Plot of the DFA3 "Fluctuation function" $F(s)$ vs. $s$ of the "length time series" for an English document (A Christmas Carol by Dickens). The length of the series is 28713 words. Again a power law behavior is observed and again the slope is equal to 0.48 very close to that of a signal with no or short range correlations. (b) Same analysis for the "Frequency time series" for the same English document. The initial slope here is 0.49, almost identical to that of an uncorrelated signal.

4. (a) A "length time series" for the Linux Kernel. (b) DFA3 for the Linux Kernel. The signal seems to exhibit long range correlations as the slope of the straight line is equal to 0.64.

5. (a) Double logarithmic plot of the correlation integral $C(r)$ for several embedding dimensionalities $n$. The exponents $d$ are calculated from the slope of the straight line segments. (b) Plot of exponent $d$ versus the dimensionalities $n$. **Rectangles**: Greek language document. **Circles**: Linux Kernel. The straight line has a slope equal to 0.69 and shows no saturation. **Triangles:** Shuffled Greek language document.

**FIGURES**

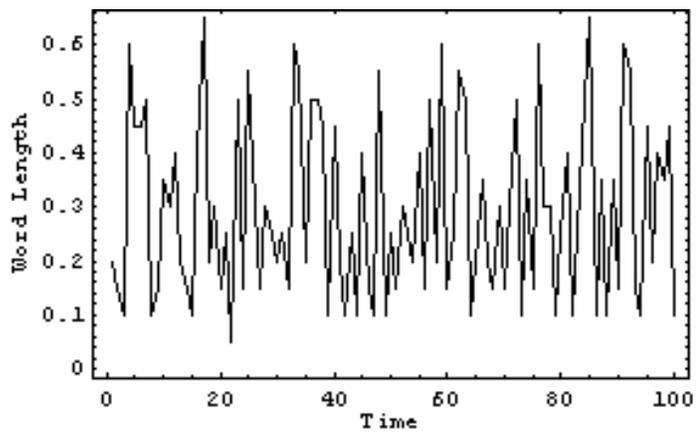

(a)

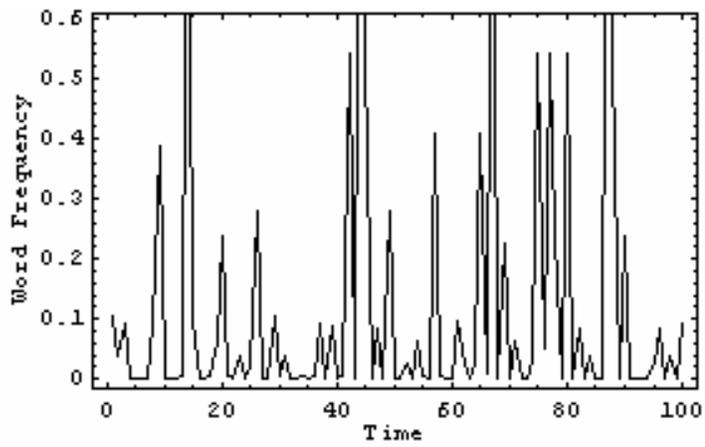

(b)

Figure 1

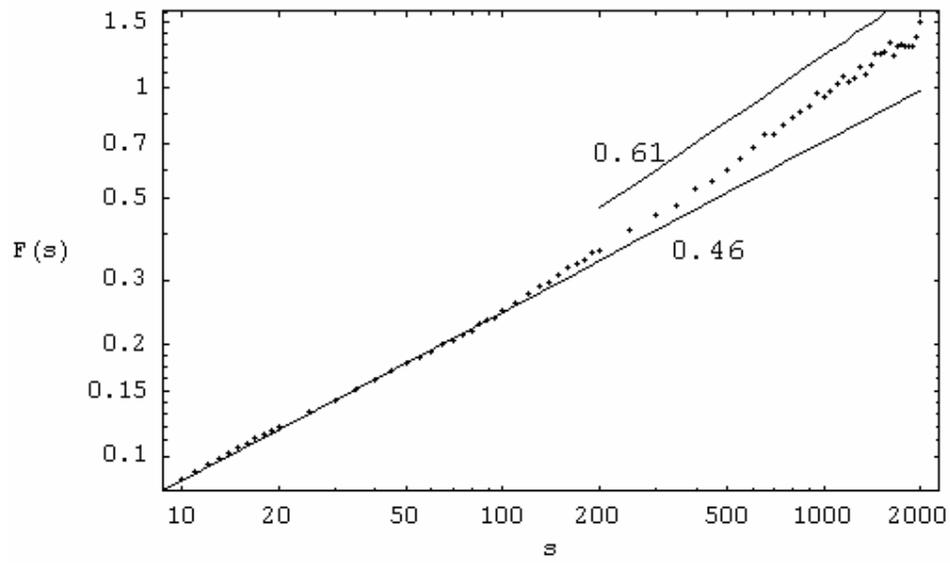

(a)

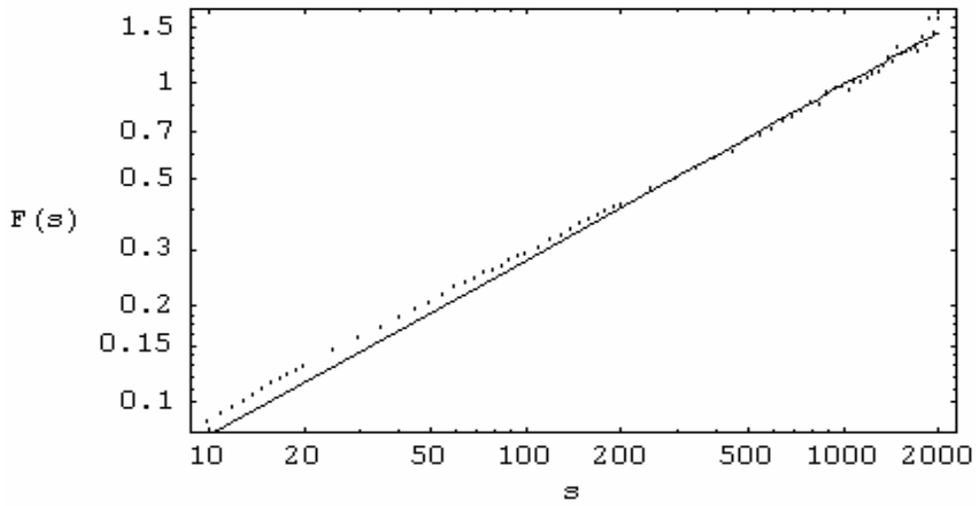

(b)

Figure 2

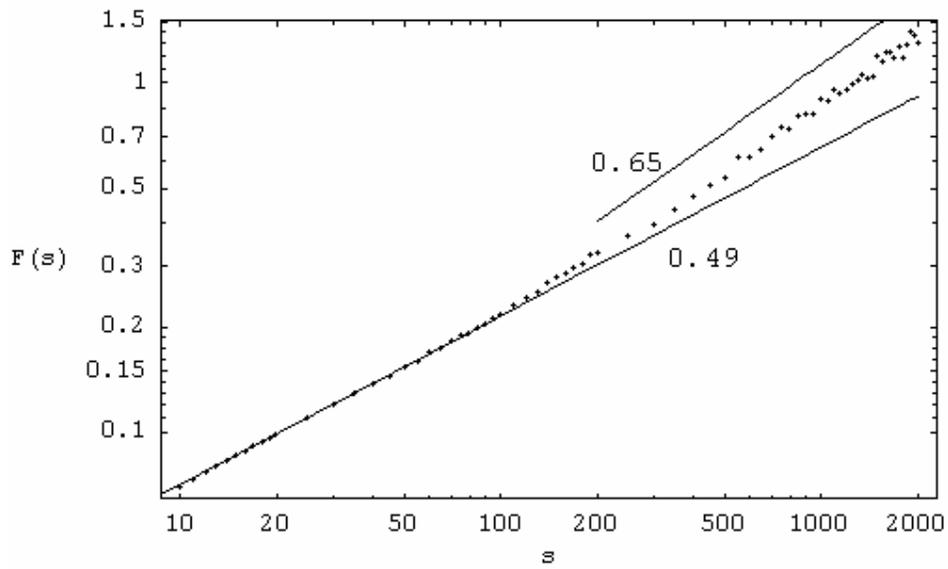

(a)

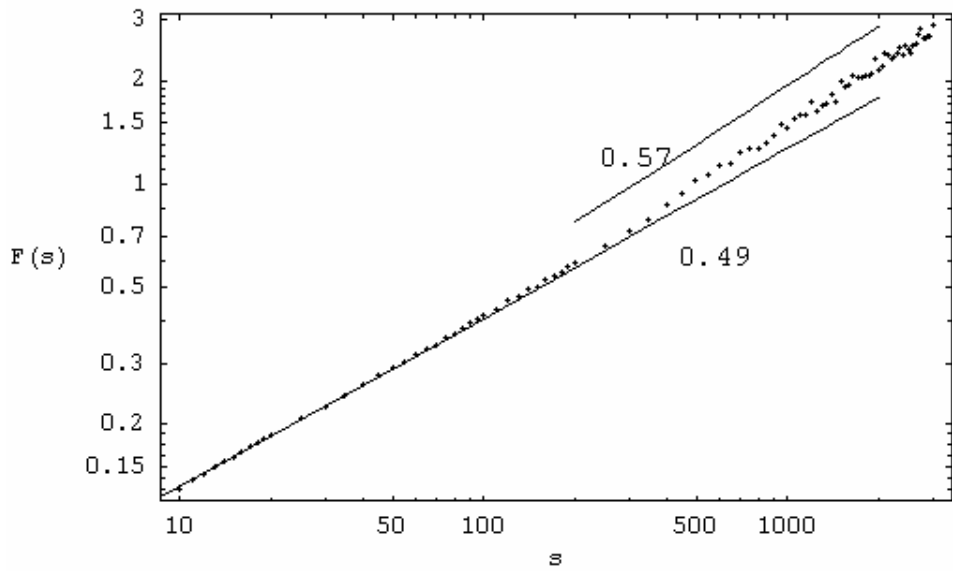

(b)

Figure 3

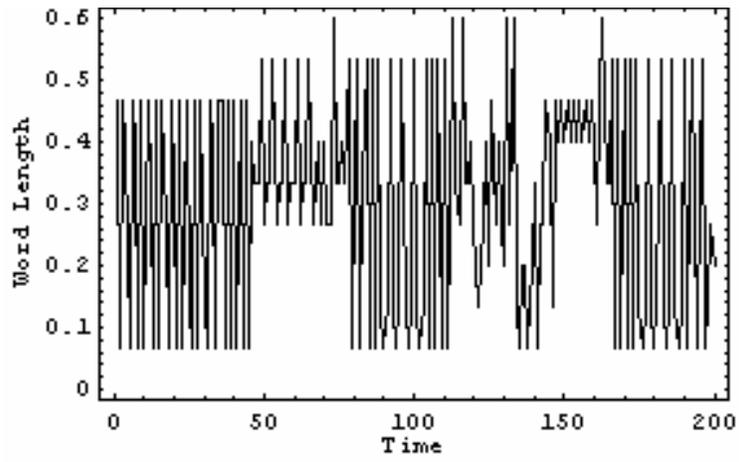

(a)

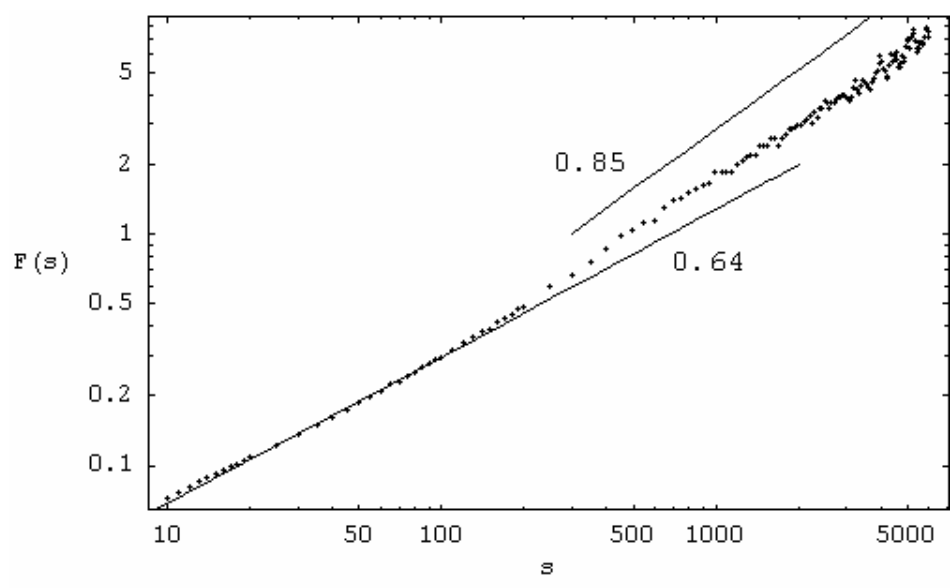

(b)

Figure 4

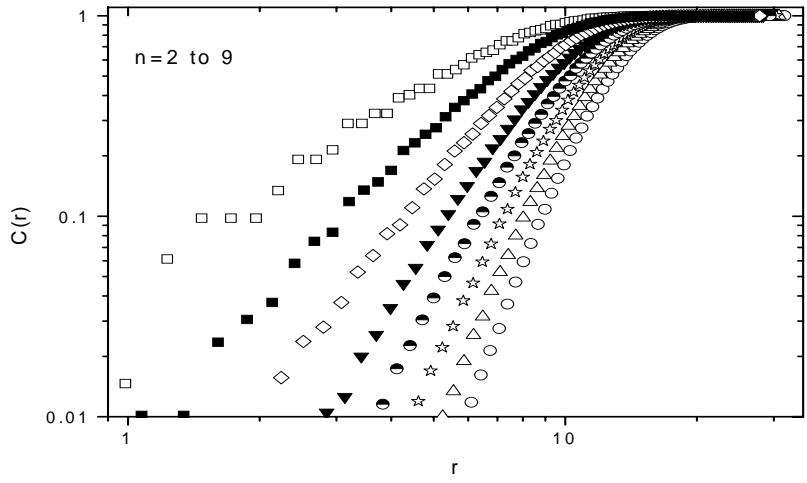

(a)

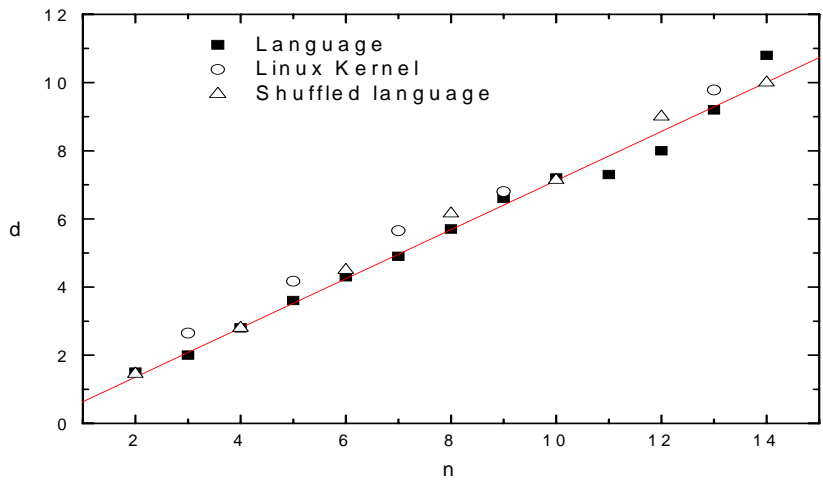

(b)

Figure 5